# Measurement of X-ray Photon Energy and Arrival Time Using a Silicon Drift Detector

Liu Li[1] (刘 利), Zheng Wei[1] (郑 伟)

[1] College of Aerospace Science and Engineering, National University of Defense Technology, Changsha 410073, China

**Abstract**: Detecting the X-ray radialization of pulsars and obtaining the photons' time of arrival are the foundation steps in autonomous navigation via X-ray pulsar measurement. The precision of a pulse's time of arrival is mainly decided by the precision of photon arrival time. In this work, a silicon drift detector is used to measure photon energy and arrival time. The measurement system consists of a signal detector, a processing unit, a signal acquisition unit, and a data receiving unit. This system acquires the energy resolution and arrival time information of photons. In particular, when background noise with different energies disturbs epoch folding, the system can acquire a high signal-to-noise ratio pulse profile. Ground test results show that this system can be applied in autonomous navigation using X-ray pulsar measurement as payload.

## 1 Introduction

X-ray pulsar navigation has been a research hotspot for several years and is a novel autonomous navigation approach to satellite running. Related papers mainly concentrate on X-ray pulsar observation and detector, navigation theory [2, 3]. A series of science programs has also been carried out for X-ray pulsar navigation [4].

According to the characteristics of X-ray pulsar navigation, the requirements for X-ray pulsar detection are as follows. First, the detection energy should range from 1 keV to 10 keV. Second, the detection requires energy resolution, which can eliminate background noise. Third, the detection time resolution should be better than or equal to 10 μs.

The common detection types are gas proportional counter, micro-channel plate (MCP), and semiconductor detector. Gas proportional counters are limited due to the gas lifetime and damage to the anode wires within the chambers [5]. Although the MCP has a high time resolution, it does not have energy resolution [6]. A silicon drift detector (SDD) is a kind of semiconductor detector. An SDD is unique and practical because it achieves simultaneous high energy resolution and fast timing compared with conventional semiconductor detectors. An SDD can also be used at near-room temperatures [7]. In an autonomous navigation-based pulsar field, an SDD can precisely provide X-ray photon energy and time information.

## 2. Structure of measurement system of photon energy and arrival time

### 2.1 Overall structure of the system

The system consists of four parts, i.e., an X-ray detector and signal processing unit, a photon energy and photon time of arrival acquisition unit, a data receiving unit, and a power supply system. In the X-ray detector and signal processing unit, the charge signal generated from the SDD detector goes through a four circuit module including a charge-sensitive amplifier, filter and shaper circuit, pole-zero cancellation, and the main amplifier, that transforms into near-Gaussian shape. In the photon energy and photon time of arrival acquisition unit, a Gaussian analog signal is transformed to a numerical signal. The energy and time information of the numerical signal is then captured by FPGA in this unit. Then, in the data receiving unit, data are transferred through the USB bus to the

Supported by National Natural Science Foundation of China (10973048).
E-mail: zhengweinudt@163.com

upper computer. In the signal processing unit, the power supply system generates +12, -12, +5, and 1.3 V. Meanwhile, the acquisition unit requires +12, -12, and +5 V.

## 2.2 SDD unit

As shown in Fig. 1, an SDD consists of fully depleted high-resistivity silicon, in which an electric field parallel to the surface and created by properly biased contiguous field strips drives signal charges toward a collecting anode. A front-end n-channel junction gate field-effect transistor is integrated onto the detector chip close to the n+ implanted anode. The extremely low anode capacitance enables higher resolution for a short shaping time. The X-ray detector efficiency is about 100% at 10 keV and 75% at 2 keV with a 25 μm-thick beryllium window [8].

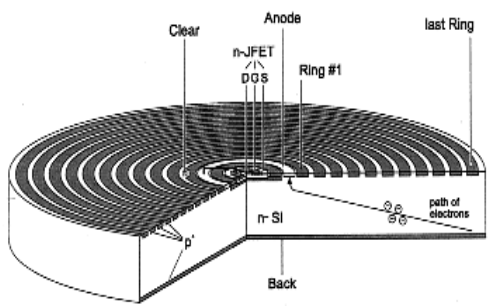

Figure 1. Cross-section of a cylindrical SDD.

## 2.3 Acquisition unit for X-ray photon energy and time of arrival

The acquisition unit for X-ray photon energy and arrival time is shown in Fig. 2. This unit contains a peak-holder circuit, event-choosing circuit, analog-to-digital conversion chip, and FPGA chip. The FPGA is the control core chip that accomplishes acquisition assignment. The main function of this chip is to receive instruction from the upper computer, control single acquisition, and send data to the upper computer through the USB bus.

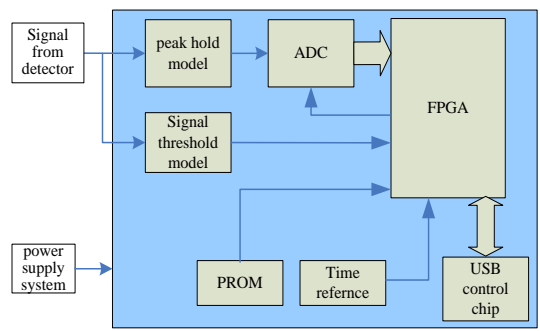

Figure 2. Electronic block diagram of the acquisition unit for photon energy and arrival time

The acquisition unit flowchart is shown in Fig. 3. The unit enables the peak hold model to capture and hold the signal's peak. If the signal amplitude is higher than the threshold of the discriminator, the ADC circuit is trigged by FPGA and starts to sample the signal's peak. The signal's peak is considered as the signal's energy.

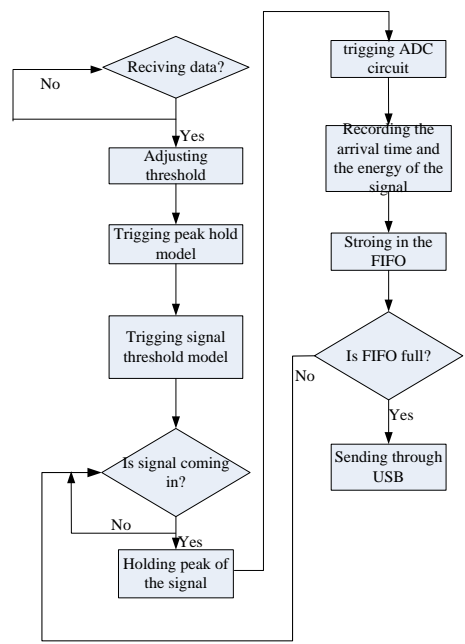

Figure 3. Flowchart of the acquisition unit.

As the ADC circuit is trigged, the signal's rising edge time is simultaneously captured by FPGA, which is recorded by the interior counter and converted to absolute time by the crystal oscillator and multiplier factor in FPGA. Energy and time information is stored in the FIFO. When the FIFO is full, the USB control model is used to send information to the upper computer and clean the FIFO. A data packet comprises 8 bytes. The first byte is the prefix, the signal's time information is in the second to sixth bytes, and energy information is in the remaining bytes.

## 3 Performance test

### 3.1 Energy resolution and energy linearity test

To determine the relationship between channels and photon energy, the energy spectrum lines of $^{55}$Fe and $^{241}$Am radioactive sources are measured to calibrate the linear energy. Based on the measurement data in Table 1, the linear formulas can be expressed as $E=0.00766C+0.1446$, where $C$ is the channel value and $E$ is the corresponding energy (in keV). The channel-to-energy linear relationship is concluded as perfect, and the correlation coefficient is >0.99.

To test the energy resolution performance of the SDD, $^{55}$Fe and $^{241}$Am source energy spectra are obtained. Fig. 4 shows that the energy resolution (FWHM) is around 155eV@ 5.90 keV and 205eV@ 13.93 keV.

Table 1. Peak energy spectrum

| Source | Channel | Energy (keV) |
|---|---|---|
| Fe$^{55}$ | 492 | 5.90 |
|  | 543 | 6.49 |
| Am$^{241}$ | 1006 | 11.87 |
|  | 1181 | 13.93 |
|  | 1503 | 17.61 |
|  | 1764 | 21.00 |

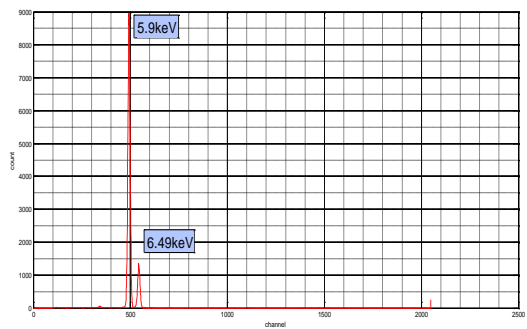

Figure 4a. Fe$^{55}$ energy spectrum (FWHM～155keV)

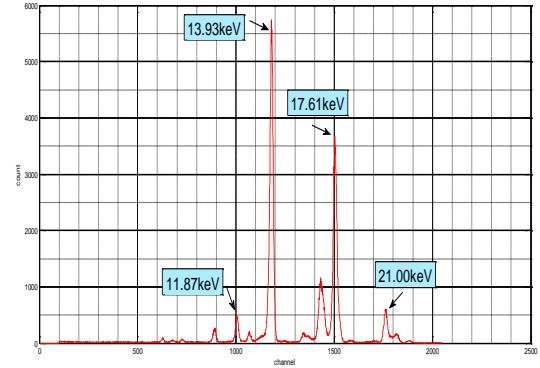

Figure 4b. Am$^{241}$ energy spectrum (FWHM～205keV)

### 3.2 Time precision and time resolution test

Time precision test

The time precision of the signal processing and acquisition units is a crucial factor affecting the time precision of the entire system. The time information of Gauss pulses is collected using the signal processing and acquisition units with a signal generator to produce Gauss pluses with 5 kHz frequency to substitute for the detector. The result of time precision is shown in Fig. 5. Most pulses are received at 200 μs (5 kHz). Only a few pulses have errors, and the RMS is 20.3 ns.

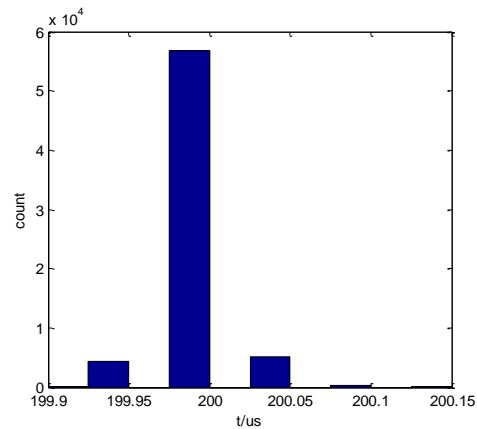

Figure 5.Time data collected by the acquisition units when the Gauss pulse is 5 kHz.

Time resolution test

The time resolution of the SDD is a significant affecting navigation application and indicates the minimum time interval between two photons received by the detector. In pulsar navigation application, the time resolution of the detector ≤ 10 μs. The statistical distribution of the time interval of two photons is presented in Fig. 6a. Fig. 6b is an amplification of Fig. 6a (top). Figs. 6a and 6b show that the time interval distribution of two photons obeys Poisson distribution. The minimum time interval of two photons is about 10 μs, namely, the detector's time resolution is 10 μs.

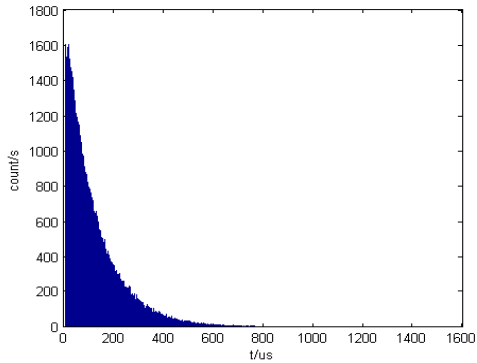

Figure 6a. Statistical distribution of time interval.

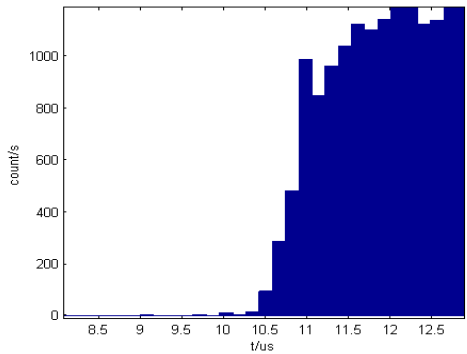

Figure 6b. System time resolution.

## 3.3 Pulse profile with background noise

Background noise is created by interaction between the detector and orbital environment particles, such as γ-ray, protons, and electrons. Background noise decreases the pulse profile SNR and precision of pulse TOA estimation. Background energy differs from the X-ray photon signal energy. Therefore, the SDD energy resolution can be used to eliminate background noise.

To validate this idea, some experiments have been developed. First, X-ray pulses with low energy and background noise with high energy are produced. Second, mixed arrival time of pulses and background noise with and without energy information, respectively, are likewise produced. Finally, energy resolution affection to the pulses SNR is analyzed.

Pulse profile with low energy

Signal simulation [9] including X-ray source and profile modulation is performed to produce periodic X-ray pulses and background noise.

The X-ray pulsar energy ranges from 1 keV to 8 keV, and Ti's energy ranges from 4 keV to 5 keV. Thus, X-rays generated from an X-ray tube bombard Ti to produce the right energy range for simulating X-ray pulsar.

When the incident energy to the system is 10 keV and the period of pulse is 25 ms, photon energy exists (Fig. 7a). Energy is mainly distributed as follows: Ti's Kα characteristic line, 4.51 keV; Ti's Kβ characteristic line, 4.51 keV; and bremsspectrum, <10 keV.

The pulse profile made by folding 3000 periods is shown in Fig. 7b. Photon flow density in the peak is about 700 counts/s. No noise photons are received by the detector, so the photon flow density of the rest is zero.

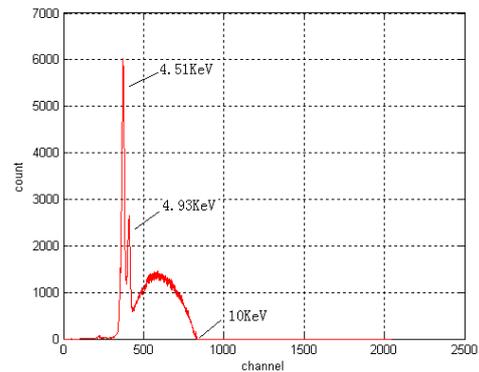

Figure 7a. Energy spectrum of pulse

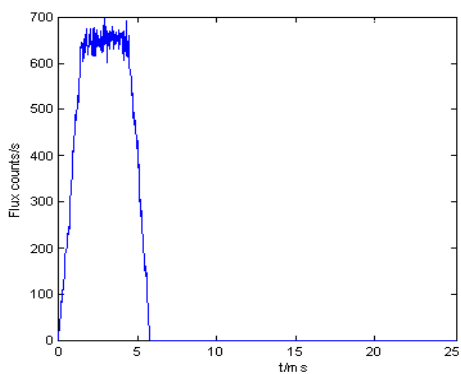

Figure 7b. Pulse profile

X-ray photon background noise

The Cu energy range is distributed within 8–9 keV. Thus, X-rays generated from an X-ray tube bombard Cu to produce the right energy range to simulate background noise.

Figure 8a shows that photon energy exists when the system's incident energy is 20 keV. Energy is mainly distributed as follows: Cu's Kα characteristic line, 8.04 keV; Cu's Kβ characteristic line, 8.9 keV; and bremsspectrum, <20 keV. Photon flow density is about 470 counts/s. Photon time information obeys normal distribution. The expectation and variance are 470 and 53, respectively.

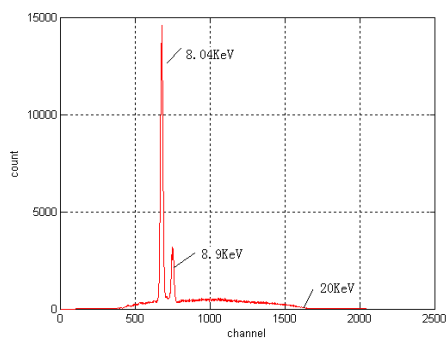

Figure 8a. Energy spectrum of noise.

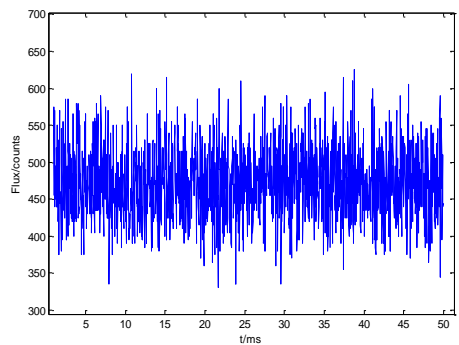

Figure 8b. Noise density of noise.

The arrival time of pulses and time information of background noise are mixed without energy information to produce a pulse profile (Fig. 9a).

SNR analysis of pulse profile with background noise

To acquire a pulse profile with high SNR, photons whose energy is in background noise area are omitted. The relationship between the deleted range and SNR is presented in Table 2. Accordingly, the SNR is higher when the deleted area is closer to the energy area of the background noise. When the deleted area is [8, 9] keV, the SNR of the pulse profile reaches the maximum of 7.25. The pulse profile is shown in Fig. 9b. The pulse profile SNR in Fig. 9a is better than that in Fig. 9b. Background noise photon energy is concentrated on the area [8, 9] keV, although only a few signal photons are located in this area. Therefore, photon removal in the area [8, 9] keV produces the best SNR of pulse profile.

Table 2. Relationship between deleted range and SNR.

| Deleted range (keV) | [5.3, 9] | [6.5, 9] | [7, 9] | [8, 9] | [8.5, 9] |
|---|---|---|---|---|---|
| SNR | 3.79 | 5.01 | 6.25 | 7.25 | 2.35 |

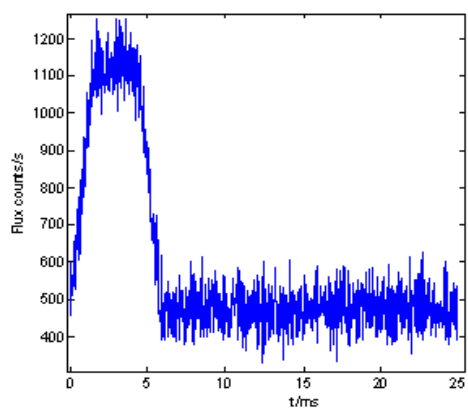

Figure 9a. Pulse profile with background noise.

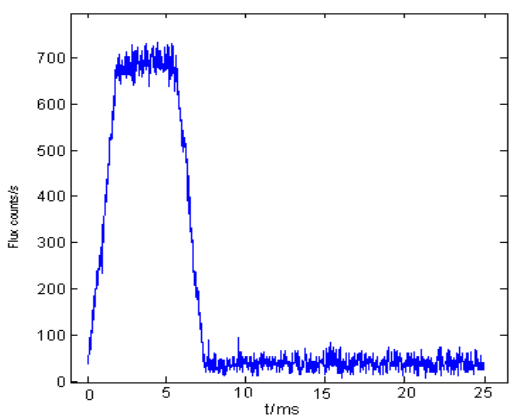

Figure 9b. Pulse profile after eliminating photons and noise in [8, 9] keV.

## Conclusions

In summary, Measurement systems for X-ray photon energy and arrival time are proposed and built. The results of the tests performance show that the system can measure the energy and time precisely and satisfy navigation requirement. We employ the energy resolution and time measurement to exclude most background noise, and acquire the pulse profile with high SNR. The measurement system can be applied in the autonomous navigation based on X-ray pulsars.